\documentclass[aps,reprint,amsmath,amssymbols,superscriptaddress]{revtex4-1}

\usepackage{graphicx}
\usepackage{bm}
\usepackage{color}
\usepackage{soul}

\newcommand{\micron}{\mu \mathit{m}}

\begin{document}


\title{Visualization, coarsening and flow dynamics of focal conic domains in simulated Smectic-A liquid crystals}


\author{Danilo B. Liarte}
\email[]{dl778@cornell.edu}
\affiliation{Institute of Physics, University of S\~ao Paulo, S\~ao Paulo, SP, Brazil}
\affiliation{Laboratory of Atomic and Solid State Physics, Cornell University, Ithaca, NY, USA}
\author{Matthew Bierbaum}
\affiliation{Laboratory of Atomic and Solid State Physics, Cornell University, Ithaca, NY, USA}
\author{Muxin Zhang}
\affiliation{Laboratory of Atomic and Solid State Physics, Cornell University, Ithaca, NY, USA}
\author{Brian D. Leahy}
\affiliation{Laboratory of Atomic and Solid State Physics, Cornell University, Ithaca, NY, USA}
\author{Itai Cohen}
\affiliation{Laboratory of Atomic and Solid State Physics, Cornell University, Ithaca, NY, USA}
\author{James P. Sethna}
\email[]{sethna@lassp.cornell.edu}
\affiliation{Laboratory of Atomic and Solid State Physics, Cornell University, Ithaca, NY, USA}


\date{\today}

\begin{abstract}
Smectic liquid crystals vividly illustrate the subtle interplay of broken
translational and orientational symmetries, by exhibiting defect structures
forming geometrically perfect confocal ellipses and hyperbolas.
Here, we develop and numerically implement an effective theory to study the dynamics of focal conic domains in smectic-A liquid crystals. We use the information about the smectic's structure and energy density provided by our simulations to develop several novel visualization tools for the focal conics. Our simulations accurately describe both simple and extensional shear, which we compare to experiments, and provide additional insight into the coarsening dynamics of focal conic domains. 
\end{abstract}


\maketitle

\section{Introduction}
\label{sec:introduction}

\begin{figure}[h]
\includegraphics[width=0.95\linewidth]{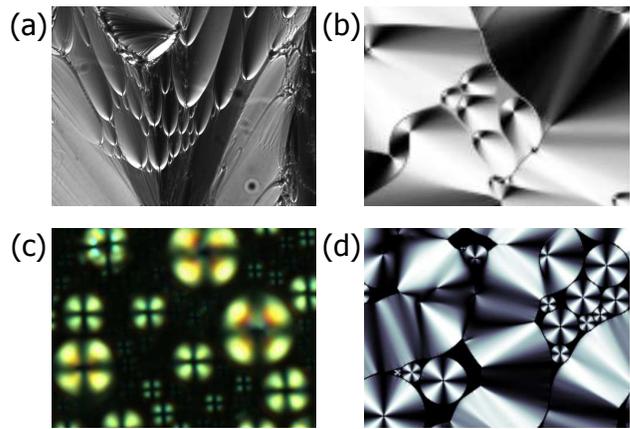}
\caption{(Color online) Experimental (a and c) and simulation (b and d) results for polarizer microscopy images of a section of smectic-A slab for planar (a and b) and homeotropic anchoring (c and d). \label{polarizers_simulation}}
\end{figure}

Translational order is {\em frail}. Most broken symmetry states
respond elastically until deformations are large. In contrast, crystals fracture or plastically yield at strains of a 
few parts per thousand. In equilibrium, they form grain boundaries -- expelling
rotation gradients into walls -- when subject to {\em atomic-scale}
boundary displacements. 
An analogous expulsion occurs in smectics, which expel
deviations from equal-layer spacing in a manner that can be mapped onto
the Meissner/Higgs effect~\cite{gennes72}. Instead of grain boundaries,
this expulsion of strain in smectics results in a remarkable patterns of
singular ellipses, hyperbolas, and parabolas known as focal conic
domains (FCDs, Fig.~\ref{polarizers_simulation}), which are the
signature of the smectic one-dimensional layered structure.
Smectics provide a window
into deep properties of translational order, lending insight into
crystalline behavior.

FCDs have appealed to theorists and experimentalists since the early
days~\cite{friedel10}, partially because of their unique geometric
origin. In its minimum energy state, a smectic has lamellar layers
spaced at equal distances. Equal layer spacing implies a singularity
at the centers of curvature of the surfaces. This constraint of equal layer spacing, surprisingly,
determines the allowed shapes of the smectic's lamella. The
lamella choose surfaces whose centers of curvature trace out
curves rather than costly two-dimensional internal
boundaries. These surfaces are called cyclides of
Dupin~\cite{hilbert99}; their centers of
curvature trace out one-dimensional conic
sections, typically confocal ellipses and hyperbolas. The resulting structures
in smectics are known as FCDs.

On a practical level, an understanding of focal conic dynamics is necessary for the description of a variety of liquid-crystalline states, such as smectic-A \cite{gennes93, kleman03}, smectic-C and C${}^*$ \cite{bourdon82, nakamura90}, lyotropic lamellar \cite{boltenhagen91, boltenhagen92}, twist-bend \cite{mandle14},  and even metallotropic liquid crystals \cite{martin06}. We focus our attention on smectic-A's, which are the simplest case. Our current understanding of focal conic structures in smectics at rest includes the study of geometrical and energetic properties \cite{sethna82, stewart94, kidney05}, the effects of anchoring for several substrates \cite{shojaei06, honglawan11, honglawan13}, the role played by dislocations \cite{meyer05, kleman06}, and beautiful insights extracted from a hidden symmetry of the Poincar\'e group \cite{alexander10}. When a smectic is driven by external dilatative stresses, experiments on initially planar-aligned samples show a sequence of elastic and plastic strain patterns that ultimately lead to a polygonal array of parabolic focal conic lines \cite{rosenblatt77}. More recently, experiments on smectic samples with antagonistic anchoring conditions subjected to shear flow report on the emergence of satellite defects \cite{anna12}. Further recent developments on the smectic rheology have been reported in~\cite{fujii10,fujii11}. However, progress in simulating smectic dynamics has been slow, perhaps because of the challenge of incorporating defect dynamics into Ericksen-Leslie-Parodi theory. Simulations of smectics are often based on atomistic and molecular dynamics approaches~\cite{glaser94,allen98,soddemann04}. Numerical solutions of the Ericksen-Leslie equations and Monte Carlo methods using the Frank free energy have been reported for nematics (see e.g.~\cite{cruz10,gruhn96}). As far as the authors know, there has been no report of the observation of focal conic domains in smectic simulations.

In this paper we present results of our simulations of an effective theory 
of smectic-A liquid crystals. Our dynamics is an extension
of Ericksen-Leslie-Parodi dynamics and the Oseen-Frank free
energy~\cite{oseen33,zocher33}, in that we allow focal conic singularites
by allowing the order parameter to change magnitude, but we continue to
forbid dislocations. The use of modern GPU computing makes these simulations feasible.
Our simulations naturally form FCDs upon relaxation of random initial
conditions and allow us to study these fascinating defects both during
formation and under mechanical loading. We find good comparisons with experiments performed under similar
situations.  Our approach allows us to investigate focal conic structures in
great detail through simulations, and provides us with an invaluable tool to understand their
several aspects, ranging from energetics, topology and geometry to anchoring
and mechanical strain effects, nicely complementing current experimental approaches \cite{Smalyukh2001}. 

\section{Equations of Motion}
\label{sec:dynamics}
Our description of the smectic starts from its elastic free energy
\begin{eqnarray}
\Psi = \int d\bm{r} \left[ F (N_\mu, \partial_\mu N_\nu) + \bm{\lambda} \cdot \nabla \times  \bm{N} \right] \quad ,
\label{total_energy}
\end{eqnarray}
which is a functional of the layer-normal field and its derivatives. The layer
normal field $\bm{N}$ can be written in terms of the scalar displacement field $u$ as $\bm{N}
= \bm{N}_0 - \nabla u$ \cite{gennes93,chaikin95}, where $\bm{N}_0$ is the undeformed layer normal. The free-energy density $F$ is given by: 
\begin{equation}
\begin{aligned}
F = \,\, &  \frac{B}{4} (1- N^4)^2 + K \, N^2 \, \left( \nabla \cdot \bm{N} \right)^2 \\
& + \frac{1}{2} K_{24} \, N^2 \, \nabla \cdot \left[\left(\bm{N} \cdot \nabla \right) \bm{N} - \bm{N} \left(\nabla \cdot \bm{N}\right)\right].
\end{aligned}
\label{energy_density}
\end{equation}
Here, the first term penalizes compression or extension of the layers away from $N=1$. The
second and third terms are related to splay and saddle-splay distortions, which
are inherited from the Oseen-Frank elastic free energy \cite{oseen33, zocher33,
frank58}. Notice that the order parameter $\bm{N}$ plays a dual role, and is very close to a unit vector field away from the focal singularities because of the small de Gennes' length (we will elaborate on this choice of dynamics in the next few paragraphs). The Lagrange multiplier $\bm{\lambda}$ forbids dislocations by ensuring that the layer-normal field is curl-free, since the vector $\nabla \times \bm{N}$ is the density of dislocations (the Burger's vector in units of the average layer spacing is given by the contour and area integrals $\oint \bm{N}\cdot d\bm{\ell}  = \int_{\Gamma} \nabla \times \bm{N} \cdot d\bm{s}$). We will treat the effects and dynamics of dislocations in a separate
paper \footnote{D. B. Liarte, A. Acharya, M. Bierbaum, and J. P. Sethna,
\textit{in preparation.}}. Note that there is no term in the free energy to account for anchoring at the boundaries. We instead enforce strict anchoring, by doubling the simulation volume and using suitably symmetrized initial conditions, to enforce the homeotropic or planar boundary conditions (see section \ref{sec:setup} for more details).

Note that a more general smectic free energy should depend on two order parameters, such as the displacement field and the tensorial liquid-crystalline order parameter. For uniaxial order, it is possible to write the elastic free energy as a functional of the Frank director $\bm{n}$ and layer normal $\bm{N}$ vector fields. Assuming these vectors are parallel (they should be in smectics-A), and that their sizes are nearly constant (they will be constant except near singularities), it is possible to minimize the free energy with respect to one of the fields, yielding a relationship between $N^2$ and $n^2$, and derive a (complicated) effective free energy in terms of a single field. For the sake of simplicity, we bypassed this analysis, and started with a single order parameter. The unusual amplitude dependence ($\sim N^2$) multiplying the $K$ and $K_{24}$ elastic terms is motivated by gradient distortions of the form $(\nabla Q)^2$, which are proportional to $N^2$ for nematic uniaxial ordering~\cite{wright89}, where $Q=((Q_{i\, j}))$ is the Maier-Saupe tensorial order parameter. Since $\bm{n}$ and $\bm{N}$ are parallel, we use $N^4$ in the first term of Eq. \eqref{energy_density} because the lowest order invariant in a Landau-de Gennes theory ($\text{tr} Q^2$) is proportional to $N^4$. Strictly speaking, neglecting an effective coupling between $\bm{n}$ and $\bm{N}$, the compression term should be proportional to $(1-N^2)^2$, as in the first term of the r.h.s. of Eq. \eqref{energy_density_II}. Later on we will get back to this choice for the smectic dynamics (see Eqs. \eqref{energy_density_II} and \eqref{energy_density_III}).

To arrive at the smectic's dynamical equations of motion, we evolve the layer normal field in the simplest possible form, assuming $\bm{N}$ relaxes directly towards equilibrium. These dynamics give a partial differential equation for the gradient-descent evolution of $\bm{N}$:
\begin{eqnarray}
\gamma \, \dot{ \bm{N} } = - \left( \frac{\delta \Psi}{ \delta \bm{N}} - \left< \frac{\delta \Psi}{ \delta \bm{N}} \right>  \right),
\label{dynamics}
\end{eqnarray}
where the angle brackets denote a spatial average and $\gamma$ is a viscosity constant; $\gamma$ can be written in terms of Leslie coefficients as $\gamma = \alpha_3 - \alpha_2$ \cite{gennes93}. The second term
of \eqref{dynamics} ensures that the net number of layers in the cell given by
$\bm{N}_0$ does not change during the gradient descent step. Equations~(\ref{total_energy}-\ref{dynamics}) differ from Ericksen-Leslie-Parodi (ELP)
dynamics in a few aspects. We relax the constraint of equal layer
spacing $|\bm{N}|=1$, which is ensured in ELP theory by means of a Lagrange
multiplier, and we consider amplitude-dependent elastic constants. Apart from the dependence on $\bm{N}$, our dynamics is a particular case of ELP theory in the limit of infinite fluid viscosity. As a result, our centers of mass move affinely with the external shear and only the orientation of the molecules change.

\begin{figure}[!ht]
\includegraphics[width=0.95\linewidth]{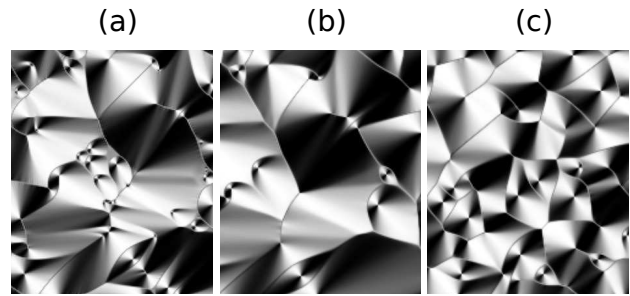}
\caption{Simulation results for polarizer microscopy images of a planar section of smectic-A slab using \emph{dynamics I} (a), \emph{II} (b), and \emph{III} (c). \label{polarizers_dynamics}}
\end{figure}

We have also considered two other choices for the energy-gradient dynamics, which are not completely described by Eqs. (\ref{total_energy}-\ref{dynamics}). For future reference, we label the dynamics described in the last paragraph as \emph{dynamics I}. For our second choice (\emph{dynamics II}), we keep equations \eqref{total_energy} and \eqref{dynamics}, but replace the free energy density by:
\begin{eqnarray}
F_{II} = \frac{B}{4} \left(1- N^2\right)^2 + K \left(\nabla \cdot \bm{N} \right)^2.
\label{energy_density_II}
\end{eqnarray}
Note that this choice of dynamics does not make contact with the tensorial order parameter $Q$. Since there is no amplitude dependence, the saddle-splay term is a surface term that vanishes in a system with periodic boundary conditions. On the one hand, the absence of a saddle-splay term limits the morphology of the allowed focal conic domains, since this term is associated with the Gaussian curvature energy of the layers~\cite{kleman03} (the splay term is associated with the mean curvature). On the other hand, the equations of motion are simpler for \emph{dynamics II}, so that we can implement simulations in a more efficient way, and study the numerical effects of varying grid sizes and de Genne's lengths (see Appendix \ref{sec:convergence}). Finally, for our third choice (\emph{dynamics III}) we consider the free energy as a functional of the displacement field and its derivatives, and replace Eq. \eqref{energy_density} by:
\begin{eqnarray}
F_{III} = \frac{B}{4} \left[1- (\nabla u)^2\right]^2 + K \left(\nabla \cdot \nabla u \right)^2,
\label{energy_density_III}
\end{eqnarray}
which is equivalent to \eqref{energy_density_II}, and Eq. \eqref{dynamics} by
\begin{eqnarray}
\frac{\gamma}{\lambda^2} \dot{u} = - \frac{\delta \Psi}{ \delta u},
\end{eqnarray}
where $\gamma$ is a viscosity constant, and $\lambda$ is a length scale that we take to be the grid spacing $a$ for convenience. This roughly corresponds to a smectic where the motion of layers is the dynamical bottleneck, rather than the reorientation of molecules (and hence the layer normals). The numerical evolution is slower for this choice of dynamics, probably due to derivatives of higher order in the equations of motion. Fig. \ref{polarizers_dynamics} shows polarizer microscopy images of a simulated smectic-A planar section, starting from the same initial condition (which has been used in Fig. \ref{polarizers_simulation}(b)) and evolved using \emph{dynamics I} (a), \emph{II} (b) and \emph{III} (c). For (a) and (b), we evolved the initial configuration for a period of about $t=2000 \tau$, where $\tau \equiv \gamma/B$. (c) was obtained using \emph{dynamics III} for a longer time ($\sim 10000 \tau$). The morphology in (c) resembles the FCD pattern shown in Fig. 2d of~\cite{martin06} for metallotropic liquid crystals.

To impose the external shear and extensional flows, we assume the layers are dragged with a displacement field determined by the flow. For simple shear, the layers are dragged in the $x$ direction according to the displacement field 
\begin{eqnarray}
u_x^{\text{s}}(x,y,z; t) = \frac{A}{l_z} \left( z - l_z \right) \sin \left( \omega t \right),
\label{displacement_shear}
\end{eqnarray}
where $l_z$ is the system size in the $z$ direction, $A$ is the amplitude, and
$\omega$ is the frequency of oscillation; our simulations are done at a fixed Ericksen number $\gamma \omega l_z^2 / K \approx 129$. 
Extensional dynamics are implemented by stretching the smectic in the $z$ direction while it contracts in the orthogonal $x$ and $y$ directions, as described by the set of equations
\begin{equation}
\begin{aligned}
l_z(t) = l_z(0) f(t), \quad l_{x,y}(t) = \frac{l_{x,y}(0)}{\sqrt{f(t)}} \quad \\
f(0)=1, \quad f(t) > 0, \, \forall \, t \in [0,\infty) \quad ,
\end{aligned}
\label{displacement_extension}
\end{equation}
where $l_x,$ $l_y$, and $l_z$ are the grid sizes along the $x,\,y,\textrm{ and }z$ directions.
To incorporate shear and dilatational dynamics simultaneously with the director relaxation, we employ an operator
splitting method, alternatively applying gradient-descent motion from
Eq.~\ref{dynamics} and one of the loading dynamics from Eqs.~\eqref{displacement_shear} \& \eqref{displacement_extension}. 

\section{Experimental and simulation setup}
\label{sec:setup}
We perform analogous experiments on 8CB in the SmA phase, using a custom-built
shear cell that allows precise control of the plate separation for gaps as
small as 2-5 $\micron$ while keeping the plates parallel to $<1$ part in $10^3$
\cite{Lin2014}, allowing us to explore a large range of strain amplitudes and Ericksen numbers. 
The shear cell is outfitted with two parallel glass plates,
which we use as the sample boundaries, and imaged with cross-polarized
microscopy. We treat the glass slides with cetyl-trimethylammonium bromide for
homeotropic anchoring and with a poly-imide treating for planar anchoring. 

At the beginning of our simulations, we generate normally distributed random
grids for each spatial component of the layer-normal field. We then enforce
anchoring constraints, and use a Gaussian filter to smooth the field on short length scales. To implement boundary conditions, we double the grid size in the z direction, and require that 
\begin{eqnarray}
&& N_x (l_z+z) = N_x (l_z-z), \, N_y (l_z+z) = N_y (l_z-z),
\nonumber \\
&& \quad N_z (l_z+z) = - N_z (l_z-z),
\end{eqnarray}
for planar anchoring, and
\begin{eqnarray}
&& N_x (l_z+z) = - N_x (l_z-z), \, N_y (l_z+z) = - N_y (l_z-z),
\nonumber \\
&& \quad N_z (l_z+z) = N_z (l_z-z),
\end{eqnarray}
for homeotropic anchoring, with $0\leq z \leq l_z - 1$. Mixed homeotropic and planar boundary conditions can be enforced in a similar way by quadrupling the thickness of the simulation grid. In order to remove the curl component of the
field, we use a Helmholtz decomposition in Fourier space. The resulting
components are divided by the mean length of the director field so that the
field has average unit norm. We use an Euler integrator with adaptive step size
\cite{press07} in order to integrate our partial differential equations. The
driving code is written in Python. Each step of the integration is evaluated
using parallel computing on a GPU using CUDA. Spatial derivatives are evaluated
with Fourier methods (FFTs).  In this letter we present results for fixed values
for the ratio $K_{24} / K = -1.5$, and for
deGenne's length scale $\xi = \sqrt{K/B} = 0.2 a$, where $a$ is the
finite-difference grid spacing. (Larger $\xi$ produces similar results with
blurred features; see Appendix \ref{sec:convergence}). Except in the study of dilatational flow, we have presented results for fixed values of $l_z=320\, \xi$. A systematic study of the dependence of the dynamical behavior on sample thickness is beyond the scope of the present work.

\section{Visualization}
\label{sec:visualization}

\begin{figure}[!ht]
\centering
\includegraphics[width=0.8\linewidth]{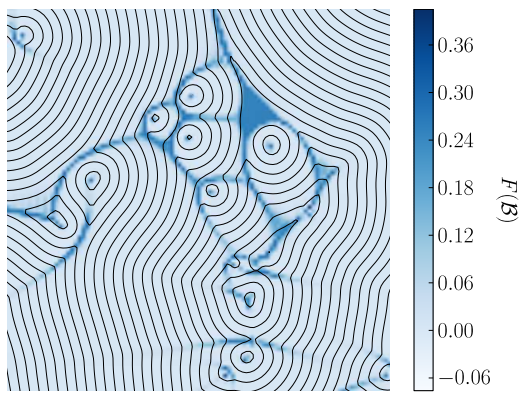}%
\caption{(Color online) Simulated energy density (white-blue density plot) and some sections of the layer surfaces (black lines) at the top section $z=l_z$ of the system with planar anchoring (Fig \ref{polarizers_simulation}b). \label{energy_density_plot}}
\end{figure}

From these random initial conditions, our gradient-descent dynamics forms 
FCDs which closely resemble those seen in experiments as shown 
in Fig.~\ref{polarizers_simulation}. We visualize the focal conics domains in our simulations through several techniques we have developed. We render polarizer microscopy images, as shown in Fig.~\ref{polarizers_simulation}, by ray-tracing light using the Fresnel equations for anisotropic dielectrics \cite{kleman03}. Figure~\ref{energy_density_plot}
shows a plot of the free energy density $F$, overlaid with cross-sections of the layer 
surfaces (contours of constant $\bm{N}_0 \cdot \bm{r} -
u$).  The free energy is high at the focal lines, where the layer contours form cusps.

\begin{figure}[!ht]
\includegraphics[width=0.95\linewidth]{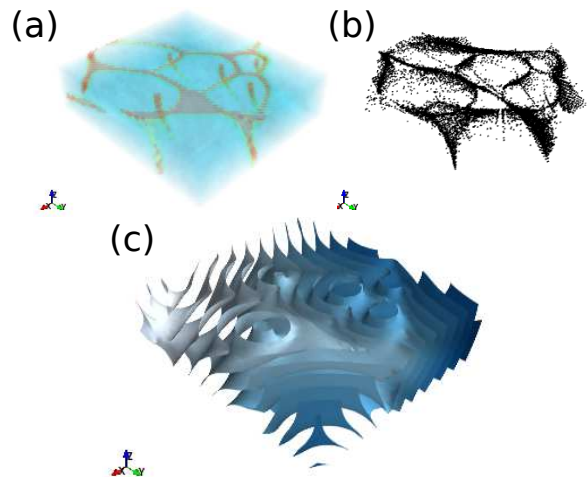}
\caption{(Color online) 3D visualizers of a simulation of smectic-A liquid crystals with planar anchoring. (a) Volume rendering visualization of the energy density; (b) loci of the centers of curvature of the layer surfaces; (c) layer surfaces.
\label{3d_visualizers}}
\end{figure}

Three visualizations of the three-dimensional smectic structure are shown in figure~\ref{3d_visualizers}.
Figure~\ref{3d_visualizers}a is a volume rendering visualization of the free energy density where each voxel is given a color and degree of transparency that is associated with its free energy density. The high energy regions (red) have organized into the characteristic focal conic structure of smectics, forming multiple ellipses, each with a hyperbola coming out of its focus. The focal conic character of the smectic structures is reinforced by the loci of the principal centers of curvature of the smectic layers, shown in Fig.~\ref{3d_visualizers}b, which coincide with the regions of high energy density in \ref{3d_visualizers}a \cite{kleman03}. To calculate the radii of curvature, we project each layer's second fundamental form tensor $M_{\mu \nu} = \partial_\mu \partial_\nu u$ into the layer-surface tangent plane. The principal radii of curvature are equal to the inverse of the principal curvatures, which are the eigenvalues of the projected $M_{\mu \nu}$. The shared surface normals intersect at the centers of curvature for the layers, which form the confocal conics \cite{sethna82}. Finally, Fig.~\ref{3d_visualizers}c shows three-dimensional level surfaces of the mass-density field.

\section{Coarsening}
\label{sec:coarsening}

\begin{figure}[!ht]
\includegraphics[width=0.9\linewidth]{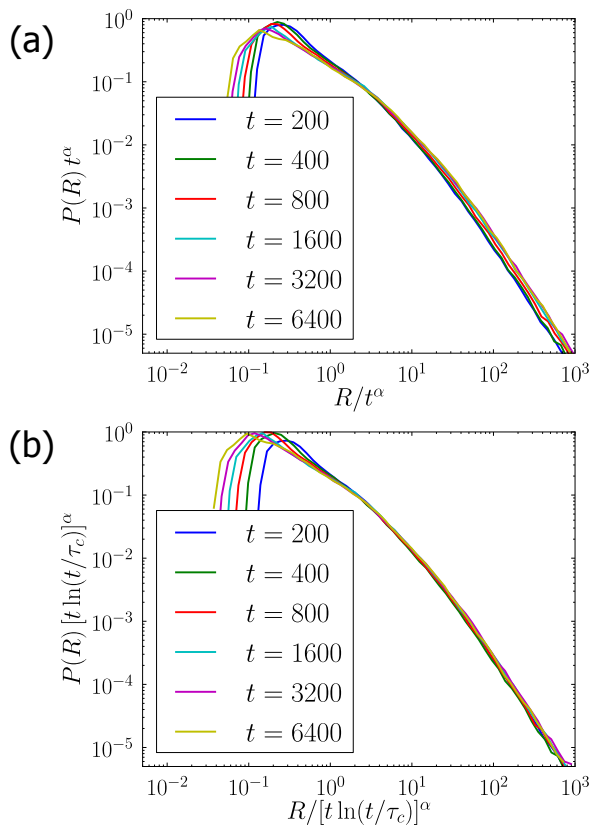}
\caption{(Color online) Scaling collapses in logarithmic scale of the layer-surface radii of curvature
distribution. (a)~Naive power-law scaling, $R^*(t) \sim t^{1/4}$.
(b)~Incorporation of logarithmic corrections to scaling, 
$R^*(t) \sim [t \log(t/\tau_c)]^{1/4}$, with $\tau_c=100\tau$.
\label{scaling}}
\end{figure}

To study the coarsening dynamics of focal conics we simulate with no anchoring at the boundary, since the boundary constraint introduces a length scale for the distribution of the layer-surface radii of curvature. As the system evolves, our dynamics seem to energetically favor ellipses with large linear eccentricity $c$. The layers around singular ellipses become flatter with increasing $c$, and converge to planes when $c \rightarrow \infty$. This is the dominant coarsening mechanism in our simulations. The coarsening of focal conics becomes slower with increasing time, but it does not stop until a uniform flat configuration is reached. (Our computational defect structures can be stabilized with simulated `dust' particles on the glass slides, by introducing spatially-dependendent energetic anchoring on the boundaries.) To quantify the coarsening, we investigate the probability distribution of the principal radii of curvature $R$, which define a characteristic distance to the focal conics, and are distributed according to a function $P(R,t)$ which also depends on time. The scaling assumption states that the morphology at late times statistically scales with a single length scale $R^*$, so in particular $P(R,t) \propto \Pi(R/R^*(t))$ for some (possibly universal) function $\Pi(X)$. In Appendix \ref{sec:scaling}, we propose two possible arguments yielding the cutoff radius of curvature $R^*\sim t^{1/4}$, and $R^* \sim [t \ln (t/\tau_c)]^{1/4}$. Fig \ref{scaling} shows scaling collapse plots that are consistent with both of these possibilities, (see Appendix \ref{sec:scaling} for more details, and for a discussion of the decay in the energy density with time). Using the first scaling form ($\sim t^{\alpha}$), we observe that the exponent $\alpha=0.5$ gives a better numerical collapse of the data. We do not show the collapse plot for this exponent because it does not have theoretical motivation. Incidentally, the inclusion of logarithmic corrections (due to the singularities near the conic sections) makes the collapse worse for $\alpha=0.5$, but improves the collapse using $\alpha=0.25$. These results do not change if we use \emph{dynamics II}~\footnote{Previously, using a different amplitude dependence of the elastic constants ($\sim N^4$), we were motivated by an apparent self-similar structure to propose a third scaling form ($\sim t^\alpha$, with $\alpha \approx 0.29$). The interested reader should refer to the second version of this manuscript in the arXiv repository~\cite{liarte15}.}. The approach to equilibrium by \textit{increasing} eccentricity to minimize bending energy is an interesting contrast to the typical approach to equilibrium of decreasing eccentricity to minimize a surface energy.

\section{Flow dynamics}
\label{sec:flow}
Our simulations and experiments also provide a window to understand dynamics of focal conic domains under shear. From our simulations, simple shear oscillations parallel to the glass slides primarily act to accelerate the focal conics' coarsening. When we shear stabilized focal conic structures, our simulations show that the focal conics are not significantly altered by the planar shear, in qualitative agreement with our experiments with strong homeotropic anchoring. In addition, our simulations allow us to tune the smectic's anchoring at the boundary. As a result, our simulations promise to discern the effects of anchoring imperfections, such as weak or spatially-modulated anchoring, on the rich structures that can be produced in experiments (see supplemental animations~\cite{url:FocalConics}). 

\begin{figure}[!ht]
\includegraphics[width=0.95\linewidth]{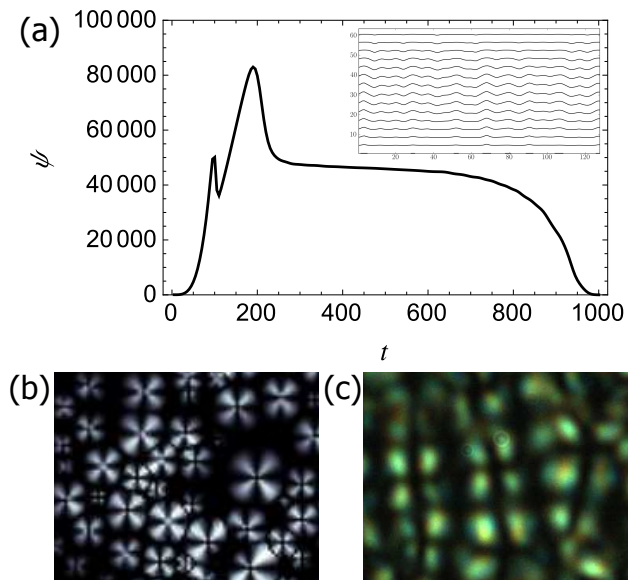}
\caption{(Color online) Simulation and experimental results for SmA under dilative stress. (a) Total free energy as a function of time. In the inset we show layers in the $x$-$z$ plane showing undulation pattern at $t =120 \tau$. (b) Simulation and (c) experimental results for crossed-polarizer images showing a pattern of focal conic domains, at strain amplitudes of 0.33 and 0.13, respectively. \label{dilation_results}}
\end{figure}

Under dilative strain (stretching the layer spacing), homeotropic smectic-A
liquid crystals are known to release free energy by forming
undulations \cite{delaye73, clark73, clark82}, and focal conic domains
\cite{rosenblatt77, Chatterjee2012}. In Fig.~\ref{dilation_results}a we show simulation results
for the total free energy as a function of time for a dilative dynamics with
$f(t) = 1 + A (1-\cos \omega t)$, where $A = 0.25$, and $\omega= 2\pi / 1000 \,
(\tau^{-1})$. The first sharp peak at about $t_1 = 100 \, \tau$ marks the onset
of an undulation pattern, which is depicted in the layers contour plot in the inset of Fig.
\ref{dilation_results}a. Linear stability analysis using the methods of
\cite{delaye73, clark73, clark82} leads to a critical strain threshold $\epsilon_c$ that is given
by the solution of the equation (see Appendix \ref{sec:stability}):
\begin{eqnarray}
\epsilon_c = \frac{\pi \xi}{{l_z}} \sqrt{ \frac{1 - 6 \epsilon_c + 6{\epsilon_c}^2}{\left(1 - 3\epsilon_c + 2 {\epsilon_c}^2\right)^2}}
\approx
\frac{\pi \xi}{{l_z}},
\end{eqnarray}
since $\epsilon_c$ is small. This analysis results in a buckling wavelength of
$\approx 9a\approx0.04 l_x$, which is consistent with our simulations (see inset of Fig.
\ref{dilation_results}a), as is the onset strain of the instability (the first peak in Fig. \ref{dilation_results}a) is later than the instability onset by approximately a factor of two, because of the growth time of the undulation
pattern). The second peak of the free energy signals the onset of a
configuration which evolves towards a complex pattern of focal conic domains.
Fig.~\ref{dilation_results}b and \ref{dilation_results}c show crossed-polarizer
images obtained from simulations and experiments at maximum strain,
respectively. We found compatible results using \emph{dynamics II}.

Note the fascinating fact that the critical change in length $\epsilon l_z
\approx \pi \xi$ is a microscopic length. Except near a critical point, one
expects $\xi$ to be of the order of a molecular size; the instability threshold~\footnote{This instability to defect structures is {\em not} true for
    two-dimensional smectics with undulating~\cite{Durand72,clark73}
    boundary conditions~\cite[Ch.~4]{Huang91}.}
for a bulk material happens when one stretches it by one molecular length \cite{Larson1999}! A simple
calculation for crystals shows an analogous result for grain boundaries: a bent
crystal's ground state has dislocations once the net displacements become of
the order of the lattice constant (up to a logarithm of the crystal size over the
atomic size). 
Unlike crystals which are metastable, smectics are unstable under long-wavelength
deformations with atomic-scale displacements -- the lower-energy defective state has no associated nucleation barrier.
Thus the equilibrium continuum elastic theory of materials with
broken translation invariance is frail \cite{Buchel1996, Buchel1997}-- it is only valid in general for
{\em microscale} net displacement differences over {\em macroscale} distances. 

\section{Conclusions}
\label{sec:conclusions}
To conclude, we have presented results from numerical simulations and experiments of smectic-A liquid crystals. In our simulations, focal conic domains spontaneously emerge out of random initial configurations. The numerical reproduction of the experimental morphologies is striking, for both planar and homeotropic boundary conditions. Our several visualization tools comprise the description of the most important structural features of FCDs. We have also characterized the coarsening of FCDs, by studying the scaling behaviour associated with the distribution of the layer surface radii of curvature, which is the length scale related to the size of the focal conics. Finally, we have studied the system response to strain, which includes a numerical and experimental investigation of the classical dilatational instability, correctly predicting the instability threshold, and the onset of a state populated with parabolic focal conic domains.

\appendix

\section{Convergence tests}
\label{sec:convergence}

In this section we present some results of a test to analyze the effects of small grid sizes and small de Genne's length ($\xi=0.2 a$) in our simulations. We start with a smoothened random initial field of linear size $16 a$, and evolve it for a very short time ($\approx 0.5 \tau$), using \emph{dynamics II} with $\xi=0.2 a$~\footnote{We do not use \emph{dynamics I} in these calculations because it is harder to control numerical instabilities for larger de Genne's length.} and no anchoring at the boundaries. We then duplicate the resulting configuration into larger lattices, with linear size $32$, $64$ and $128$. To generate a smooth interpolation between lattice points of the larger lattices, we copy the Fourier components of small wave number, and leave the coefficients associated with short wavelengths equal zero. To be consistent with the periodic boundary conditions, we divide the smaller cubic grid into eight equal pieces, and copy the configuration of each piece into the corresponding corner of the larger grid in Fourier space. To estimate the finite-size error, we evaluate
\begin{eqnarray}
\sigma_b = \max_{\mu \in \{x,y,z\}} \max_{i,j,k} \left| N_\mu(b \, i, b \, j, b \, k) - N^\prime_{\mu}(i,j,k) ) \right|
\end{eqnarray}
where $\bm{N}$ and $\bm{N}^\prime$ are the large and small lattices, with linear sizes $l$ and $l^\prime = b \, l$, respectively. The indices $i$, $j$, and $k$ are grid coordinates of $\bm{N}^\prime$. The second row of Table \ref{error_doubling} shows the error comparison for this initial configuration. Since the configuration is copied (with a smooth interpolation) from the smaller to the larger lattices, this error is of order $10^{-15}$. We then evolve this initial state for each grid for a period of time of about $200 \tau$, keeping $\xi=0.2a$ for $l = 16$, and using \emph{dynamics II}. In order to have comparable simulations, we multiply $K$ and $K_{24}$ by four (thus increasing $\xi$ by a factor of two) each time we double the grid size, since the wave vectors are divided by two, and the splay and saddle-splay terms contribute with two gradient terms. The results for $\xi_{L=16}=0.2a$ and $t=200\tau$ are shown in the third row of Table \ref{error_doubling}. Note that the difference between the $32^3$ lattice (with $\xi=0.4a$) and the $128^3$ lattice (with $\xi=1.6a$) is of just about two percent. For the sake of completeness, we started with the same initial state, and evolved each lattice using \emph{dynamics II} with $\xi_{L=16}=0.4a$ (so that $\xi=0.8a,1.6a$, and $3.2a$ for $L=32,64$ and $128$, respectively). The results for this case are shown in the fourth row. Notice that $\sigma_2(64,128)$ for $\xi_{L=16}=0.2a$ is comparable to $\sigma_4(32,128)$ for $\xi_{L=16}=0.4a$, because the $64^3$ lattice in the third row and the $32^3$ lattice in the fourth row are simulated using the same de Genne's length. The same comparison holds between $\sigma_2(32,128)$ for $\xi_{L=16}=0.2a$ and $\sigma_4(16,128)$ for $\xi_{L=16}=0.4a$. Note that this test analyzes convergence upon increasing both resolution and de Genne's length, and that $\xi=0.4 a$ gives sensible results even for small lattices. We recall that most of the results that have been presented in this paper were obtained using \emph{dynamics I} with $\xi=0.2a$ for large grids ($256^2 \times 64$). We keep the results for \emph{dynamics I}, even though it is harder to control numerical instabilities in this case, because that is our only choice with a Gaussian curvature energy, which is associated with the saddle-splay term. We emphasize that we do not observe a significant change for our results for coarsening and dilative dynamics when we use \emph{dynamics II} with $\xi=0.4$.

\begin{table}[h]
\centering
\begin{tabular}{ | l | c c c | }
\hline
& $\sigma_8$ $(16, 128)$ &  $\sigma_4$ $(32, 128)$ &  $\sigma_2$ $(64, 128)$ \\ \hline
Initial state & $10^{-15}$ & $10^{-15}$ & $10^{-15}$  \\
$\xi_{L=16}=0.2a$ at $t = 200\tau$ & $0.55$ & $0.03$ & $0.0007$  \\
$\xi_{L=16}=0.4a$ at $t = 200\tau$ & $0.02$ & $0.0009$ & $0.0007$  \\
\hline
\end{tabular}
\footnotesize
\caption{ Comparison of the errors of evolved simulation configurations for grids with increasing resolution and de Genne's length. }
\label{error_doubling}
\end{table}

\section{Scaling exponent for the coarsening of focal conics}
\label{sec:scaling}

The principal radii of curvature $R$ define a characteristic distance to
the focal conics: equally spaced layers develop singularities at their centers of curvature. These radii have a time-dependent probability distribution $P(R|t)$. Scaling suggests that all correlation functions should scale with a single length scale $R^*(t)$ that
diverges at late times, hence $P(R|t) \approx \Pi(R/R^*) / R^*$ for some
perhaps universal function $\Pi(X)$. (Here
the last factor preserves normalization: $\int P(R) dR = \int \Pi(X) dX = 1$.) 
In coarsening problems, it is often possible to use simple energetic
arguments to derive the power law divergence $R^*(t) \propto t^\alpha$;
for example, phase separation in systems without hydrodynamic flow has
$\alpha = 1/3$ for conserved order parameters and $\alpha = 1/2$ for
non-conserved order parameters. Here we give two possible scaling
forms, of increasing sophistication. The first mimics the standard energetic arguments and the second
provides a refined argument including a logarithmic correction due
to defect cores (Fig.~\ref{scaling}).

\newcommand{\EnDen}{{\cal{E}}}

\begin{figure}
\begin{center}
\includegraphics[width=\linewidth]{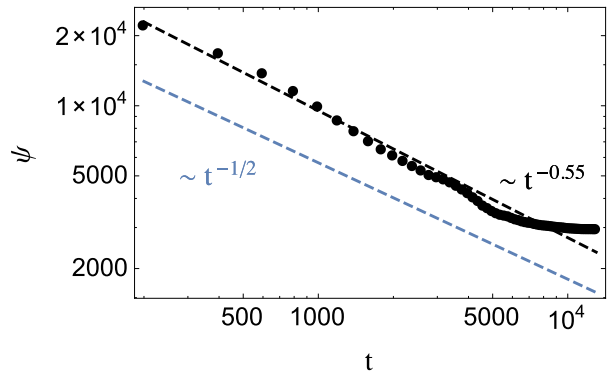}%
\end{center}
\caption{(Color online) Total free energy $\psi$ (symbols) as a function of time. The blue dashed line correspond to the behavior predicted from the
naive argument $R^*\sim t^{1/4}$. The black dashed line is a best fit.
\label{energy_decay}}
\end{figure}

Away from the defect cores, where $|\bm{N}| \approx 1$, the free energy density scales as $R^{-2}$. So the average energy density is 
\begin{equation}
\EnDen(R^*) = \int P(R) /R^2 dR.
\end{equation}
In traditional coarsening, one assumes that the integral for $\EnDen(R^*)$
converges at zero, so $\EnDen(R^*) \sim 1/(R^*)^2$. This leads to a force
(tension) $T =\delta \EnDen/\delta R^* \sim 1/(R^*)^3$. Since the order
parameter is non-conserved~\footnote{\emph{Dynamics III} could perhaps correspond to conserved dynamics.},
\begin{equation}
\dot{R} = -\gamma\, T,
\label{coarsening}
\end{equation}
where $\gamma$ is an effective inverse viscosity (see section 11.4 of
\cite{sethna06}). This can be solved giving $R^* \sim t^{1/4}$
(Fig.~\ref{scaling}a), and hence $\EnDen(R^*(t)) \sim 1/t^{1/2}$ 
(Fig.~\ref{energy_decay}).

How does this change if we consider the defect cores, where $|\bm{N}|<1$?
The energy in the cones, near the focal conic line singularities, scales as
the length of the conics times $\ln (R/\xi)$, where $\xi$ is de Gennes' length
scale. Within a focal domain of size $R^*$, near the singular ellipse and hyperbola $R\rightarrow 0$, the volume fraction $P(R) \sim R$, 
so that $\Pi(X) \sim X$ for small $X$. This leads to a divergence in the integrated
energy near the focal conic singularities, which is cut off by $\xi$,
\begin{eqnarray}
\EnDen(R^*) &=& \int_\xi^\infty (1/R^2) \Pi(R/R^*) / R^* dR
\nonumber \\ &=& \int_{\xi/R^*}^\infty \Pi(X)/(X R^*)^2) dX \nonumber \\
&\sim& (1/R^*)^2 \int_{\xi/R^*}^1 X/X^2 dX \nonumber \\ 
&=& \log(\xi/R^*) / (R^*)^2,
\end{eqnarray}
(see section 10.5 of ~\cite{kleman03}).
After some calculation, Eq.~\eqref{coarsening} implies
  \begin{equation}
  R^* \sim \left[ t \ln (t /\tau_c) \right]^{1/4},
  \label{eq:log_correction}
  \end{equation}
in the limit of large $R$ or $t$. So if the 
focal domains are all of the same length scale $R^*$, and the relaxation
of the core singularites dominates the coarsening, we expect a $t^{1/4}$
scaling with a logarithmic correction, as in Fig.~\ref{scaling}(b).
There is a large range of values for $\tau_c$ which collapse our data.
Fig.~\ref{scaling}b shows a scaling collapse plot with the logarithmic
corrections for $\tau_c=100\tau$. Note that Eq. \eqref{eq:log_correction} only applies for times times $t$ larger than $\tau_c$, hence the range of times used in the collapse plot of Fig.~\ref{scaling}(b). Our data for $t< 200\tau$ do not fit well in the collapse plot, even when we consider lower $\tau_c$ so that $t$ is still greater than $\tau_c$; we surmise that $\tau_c$ is associated with the time needed to form line singularities. Unfortunately, we have not been able to verify this hypothesis, since there is no surface anchoring in this case and the three-dimensional visualizers are not useful at early stages of the dynamic evolution.

Both of these scaling forms are compatible with the data, given the limited
scaling regime (less than a decade in length, corresponding to less than
three decades in `size'); $P(R,t)$ is clearly still evolving in shape from its non-universal initial form.

\section{Linear stability analysis for SmA under dilative strain}
\label{sec:stability}

We consider a situation where a thin slab of homeotropic smectic-A is subject to dilative stress \cite{delaye73, clark73, clark82, gennes93, chaikin95, kleman03}. In this case, the smectic layers are parallel to the glass slides, so that the stretching of the gap promotes an increase of the interlayer spacing. Planar-layer configurations store a considerable amount of bulk energy as strain is increased, which is released with the formation of an undulation pattern after a critical strain is reached. Here we use the methods of \cite{delaye73, clark73, clark82} to study the formation of undulation instabilities on smectic-A liquid crystals.

\begin{figure}
\begin{center}
\includegraphics[width=\linewidth]{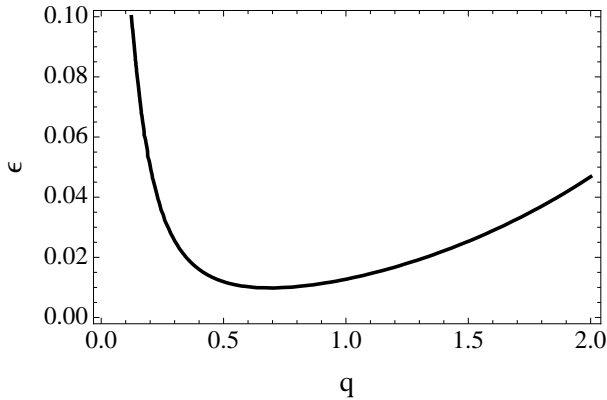}%
\end{center}
\caption{Showing critical strain as a function of $q$ for $\xi = 0.2 a$ and $l_z = 64 a$. \label{strain_q}}
\end{figure}

The displacement field associated with an undulation pattern of a smectic-A  can be written as:
\begin{eqnarray}
u(\bm{r}) = \epsilon z + u_0 \cos (qx) \sin (kz),
\label{displacement}
\end{eqnarray}
where we take $k = \pi / l_z$ to enforce strict homeotropic anchoring. Our elastic free energy density is given by:
\begin{eqnarray}
f &=&  \frac{ B }{4} (1- N^4)^2 + K \left( \nabla \cdot \bm{N} \right)^2.
\label{free_energy_N}
\end{eqnarray}
Notice that we have not included a saddle-splay term, nor have we considered amplitude dependence of the elastic constant $K$, since their effects are negligible. Roughly speaking, the amplitude dependence of $K_{24}$ gives rise to higher-order terms for the undulating solution. Hence we can approximate $N^2$ multiplying $K$ and $K_{24}$ by one, and the saddle-splay term becomes a surface term that vanishes for periodic boundary conditions. Also, we do not need include a Lagrange multiplier, since $\bm{N} = \bm{z} - \nabla u$ is curl-less if $u$ is given by Eq. \eqref{displacement}. The free energy density \eqref{free_energy_N} can be written in terms of the displacement field as
\begin{eqnarray}
f &=& \frac{B}{4} \left\{ 1 - \left[ 1 + \left( \frac{ \partial u }{ \partial x } \right)^2 - 2 \, \frac{\partial u}{ \partial z } \right]^2 \right\}^2 
\nonumber \\ && \quad
+ K \left( \frac{\partial^2 u}{\partial x^2} \right)^2.
\label{free_energy_u}
\end{eqnarray}
We can combine Eqs. \eqref{free_energy_u} and \eqref{displacement} in order to write
\begin{eqnarray}
\frac{f}{B} &=& \xi ^2 q^4 {u_0}^2 \sin ^2(k z) \cos ^2(q x)
\nonumber \\ && \quad +\frac{1}{4} \left\{ \left[ q^2 {u_0}^2 \sin^2(k z) \sin ^2(q x)
\right. \right. \nonumber \\
&& \quad \left. \left. - 2 (k {u_0} \cos (k z) \cos (q x)+\epsilon )+1\right]^2-1\right\}^2,
\end{eqnarray}
where $\xi = \sqrt{K/B}$ is de Gennes' length scale. To find the stability threshold we integrate out the free energy density over one period in the $x$-direction, and from $0$ to $l_z$ in the $z$-direction:
\begin{eqnarray}
F = \int_{0}^{\frac{2\pi}{q}} dx \int_{0}^{l_z} dz \, f(x,z).
\end{eqnarray}
The stability threshold is given by the solution of the equation:
\begin{eqnarray}
\frac{\partial^2 F}{\partial {u_0}^2} = 0,
\end{eqnarray}
or,
\begin{eqnarray}
&& 4 k^2 \left(6 \epsilon ^2-6 \epsilon +1\right)
\nonumber \\ && \quad 
+q^2 \left(\xi ^2 q^2-8 \epsilon ^3+12 \epsilon^2-4 \epsilon \right) = 0.
\label{critical_strain_1}
\end{eqnarray}
For given $\xi$ and $l_z$, this equation defines a curve in the $\epsilon \times q$ plane. Fig \ref{strain_q} shows the critical strain as a function of $q$ for $\xi = 0.2 a$ and $l_z = 64 a$, corresponding to our simulation parameters, where $a$ is the finite-difference grid spacing. The strain is minimal for
\begin{eqnarray}
q = \frac{\sqrt{4 \epsilon ^3-6 \epsilon ^2+2 \epsilon }}{\xi }.
\label{wavenumber}
\end{eqnarray}
Eq. \eqref{wavenumber} can be plugged back into Eq. \eqref{critical_strain_1}, so that,
\begin{eqnarray}
\epsilon_c = \frac{\pi \xi}{{l_z}} \sqrt{ \frac{1 - 6 \epsilon_c + 6{\epsilon_c}^2}{\left(1 - 3\epsilon_c + 2 {\epsilon_c}^2\right)^2}}
\approx
\frac{\pi \xi}{{l_z}},
\label{strain_threshold}
\end{eqnarray}
where the approximate solution on the r.h.s. of \eqref{strain_threshold} is valid since $\epsilon$ is small. Notice that our approximate critical strain ($\pi \xi / l_z$) corresponds to half of the value obtained in \cite{delaye73, clark73, clark82}, because we use a slightly different form for the free energy density. Also, it is interesting to point out that the critical change in length $\epsilon_c l_z \approx \pi \xi$ is a microscopic length (see main text). 

Fig \ref{dilation_results} shows simulation results for smectic-A liquid crystals under dilative stress with $\xi=0.2 a$ and $l_z = 64 a$. We simulate $f(t) = 1 + A (1-\cos (\omega t))$ (see Eq. (6) of main text), with $A = 0.25$, and $\omega = 2\pi / 1000 \, \tau^{-1}$. Eq \eqref{strain_threshold} predicts a strain threshold of $\epsilon_c \approx 0.01$, and a critical wavelength of $q \approx 9 a \approx 0.04 l_x$, which is consistent with our simulations (Fig \ref{dilation_results}b). However, since $\epsilon = 1 - f^{-1}$, we can rearrange $f(t)$ in order to write
\begin{eqnarray}
t_c = \frac{1}{\omega} \cos^{-1}\left(1 - \frac{\epsilon_c}{A(1-\epsilon_c)}\right) \approx 45 \, \tau,
\end{eqnarray}
which is about half of the time threshold associated with the first peak of Fig \ref{dilation_results}a. We suggest that the time scale associated with the growth of the undulation pattern accounts for the discrepancy between the simulation threshold and the analytical estimate. We tested our stability analysis directly by adding a small perturbation $\delta \bm{N}$ to $\bm{N}_0$. Under a gradient descent infinitesimal evolution of $\bm{N}$, we expect that $\mathcal{F} [ \delta N_z(t+\delta t) ] = \exp (\lambda_k \delta t) \mathcal{F} [ \delta N_z(t) ] $, where $\mathcal{F}$ denotes a Fourier transform operator. An exponent $\lambda_k$ is less than zero for stable planar configurations, and reaches zero at the critical strain for some wavenumber $q$. Careful numerical calculations for $\lambda_k$ lead to $t_c \approx 45 \tau$, and $q\approx 9a$, in agreement with our analytical estimate. There is no significant change (apart from shifts of numerical values) in the analysis and numerical results using \emph{dynamics II}.

\begin{acknowledgments}
We would like to thank D. Beller, R. Kamien, T. Lubensky, A. Acharya, and N. Clark for useful conversations. DBL acknowledges the financial support provided by the Brazilian agency Fapesp. This work was supported in part by the National Science Foundation CBET-PMP Award No. 1232666 (BL, MZ) and ACS PRF No. 56046-ND7 (IC), the National Defense Science and Engineeering Graduate (NDSEG) Fellowship 32 CFR 168a (BL), and the Department of Energy DOE-BES DE-FG02-07ER46393 (JPS, MKB, and DBL). This work made use of the Cornell Center for Materials Research Shared Facilities which are supported through the NSF MRSEC program (DMR-1120296).
\end{acknowledgments}

%

\end{document}